\documentclass[final,times,5p,twocolumn]{elsarticle}

\usepackage{amsmath}

\begin{document}

\title{Spontaneous magnetization of quark matter in the inhomogeneous chiral phase}
\author{R. Yoshiike, K. Nishiyama, T. Tatsumi}
\address{Department of physics, Kyoto University, Kyoto 606-8502, Japan}

\begin{abstract}
%Recently, a new phase which appears around the conventional 1st order chiral transition line in QCD phase diagram has been found in some effective models. In this phase, chiral symmetry is broken with the periodically modulating chiral condensate. 
Considering the density wave of scalar and pseudoscalar condensates, we study the response of quark matter to a weak external magnetic field.
In an external magnetic field, the energy spectrum of the lowest Landau level becomes asymmetric about zero, 
which is closely related to chiral anomaly, and gives rise to the spontaneous magnetization.
This mechanism may be one of candidates for the origin of the strong magnetic field in pulsars and/or magnetars.
\end{abstract}


\maketitle

Recently, the existence of the inhomogeneous chiral phase in the QCD phase diagram has been discussed by the analysis of the effective models such as Nambu-Jona-Lasinio (NJL) model \cite{nakano,nickel,ggl} or the Schwinger-Dyson approach \cite{muller}.
In this phase, the quark condensates spatially modulate and it is very similar to the FFLO state in superconductor \cite{ff,lo} or spin/charge density wave \cite{peierls,overhauser}.
%The existence of this phase has been discussed in the various situations, e.g. in the magnetic field \cite{review}. \par
Here, we consider ``dual chiral density wave (DCDW)'' \cite{nakano} among many kinds of form of the condensates: the quark condensates then take the form,
\begin{align}
 \Delta({\bf r}) \equiv \langle \bar{\psi} \psi \rangle + i \langle \bar{\psi} i\gamma^5 \tau_3 \psi \rangle = \Delta e^{iqz}, \label{condensate}
\end{align}
within the two-flavor QCD.
This configuration is also obtained by embedding one of the Hartree-Fock solutions in the NJL$_2$ model, so-called chiral spiral \cite{basar,basar2}.
Since the DCDW phase has been expected to appear in the moderate density region \cite{nakano}, it may be plausible that this phase is realized in neutron stars.

%In this phase,  through spectral asymmetry \cite{tatsumi}.\par
The effect of the magnetic field has been first discussed by Frolov et al. for the DCDW phase \cite{frolov}.
They have found that the spatial direction of the wavevector $\bf q$ is favored 
to be parallel to the magnetic field, and the domain of the DCDW phase is much extended in the QCD phase diagram.
In ref.~\cite{tatsumi} 
these features arises from some topological effect 
through spectral asymmetry of the quark energy; quarks exhibit an interesting feature in the presence of the magnetic field
and the energy spectrum becomes asymmetric about zero. There also appear new terms in the generalized Ginzburg-Landau expansion 
due to spectral asymmetry, which signals the novel Lifshitz point in the QCD phase diagram.
Thus, they emphasized the peculiar role of the phase degree of freedom of $\Delta({\bf r})$.

Here we further inquire this issue.
We study magnetic properties of the DCDW phase to reveal another aspect, {\it spontaneous magnetization} in the DCDW phase, which suggests a microscopic origin of the strong magnetic field in compact stars.

The origin of the strong magnetic field in compact stars has been one of the long-standing problems.
In particular, magnetars have the huge magnetic field $\sim 10^{15}{\rm G}$ on the surface \cite{thompson,olausen}.
As a candidate of the origin, amplification of the magnetic field by the dynamo mechanism, magnetorotational instability or the hypothesis of the fossil magnetic field has been proposed so far from the macroscopic point of view. Although numerical simulations have been actively performed, no definite conclusions have been obtained.
From the microscopic point of view, it has been proposed that the spontaneous magnetization emerges by spin alignment of quarks on the analogy of the electron gas \cite{tatsumi2}.
However, this phase should be developed in the low density region.
As another mechanism, it has been proposed that axial anomaly acting on the parallel layer of the pion domain wall produces magnetization in nuclear matter \cite{son,eto}.
%Furthermore, the ferromagnetic phase with the DCDW like pion condensate in nuclear matter is favored compared to the alternating structure phase due to axial anomaly and axial vector meson condensate \cite{hashimoto}. 
%In the following, we analyze the response of the quark matter in DCDW phase to the external magnetic field and discuss the possibility of spontaneous magnetization taking into account some non-perturbatibe effect of QCD, which may be a candidate of the strong magnetic field in magnetars. 
%It gives another mechanism from the microscopic point of view.
 
We use the two-flavor NJL model in the mean field approximation.
%\begin{align}
% {\cal{L}}_{\rm MF} =& \bar\psi \left\{i\Slash{D} +2G\left[\langle \bar{\psi} \psi \rangle + i\gamma^5\tau_3 \langle \bar{\psi} i\gamma^5 \tau_3 \psi \rangle\right]\right\} \psi  \notag \\
%                           &~~~~+ G\bigl[\langle\bar\psi\psi\rangle^2 + \langle\bar\psi i\gamma^5\tau_3\psi\rangle^2\bigr] \label{njl}
%\end{align}
%with the covariant derivative $D^{\mu}=\partial^{\mu}+iQA^{\mu}$, where $Q$ is the electric charge matrix in the flavor space, the chiral limit is taken and 
%we take .
%The mean fields take the form of the DCDW condensate (\ref{condensate}).
It is sufficient to consider the each flavor case because Lagrangian is diagonalized about the flavors.
Then, the Dirac Hamiltonian takes the form,
\begin{align}
 H = -i{\bf \alpha} \cdot {\bf D} - 2G\gamma^0 \left[\frac{1+\gamma_5}{2}\Delta({\bf r}) + \frac{1-\gamma_5}{2}\Delta^*({\bf r})\right] \label{ham}
\end{align}
with the covariant derivative, ${\bf D} = \nabla + ie_f{\bf A}$.
Taking the external magnetic field $\bf{B}$ along the $z$ axis, the energy spectrum constitutes the Landau levels \cite{frolov},
\begin{align}
 &E_{p_zn\zeta\epsilon}^{f} = \epsilon\sqrt {\left(\zeta\sqrt{p_z^2+m^2}+q/2 \right)^2 + 2|e_fB|n},~~~(n\neq 0) \\
 &E_{p_z\epsilon} = \epsilon\sqrt{p_z^2+m^2}+q/2,~~~(n=0) \label{energy}
\end{align}
with $\zeta=\pm1$, which denotes the spin polarization.
%is spin quantum number and $\epsilon=\pm1$, where,
%\begin{align}
 %&\omega_{\pm1} \equiv \pm\sqrt{p_z^2+m^2}+q/2. \notag
%\end{align}
%, which represents the particle or anti-particle state in $m\geq q/2$ and $m\equiv -2G\Delta$.
For $m\leq q/2$, the sign of $\epsilon$ does not always correspond to the particle or anti-particle state because the lower branch of the lowest Landau level (LLL) ($n=0$) is not always negative.
In the higher Landau levels (hLLs) ($n\neq 0$), there are four energy branches. 
On the other hand, LLL has only two energy branches and becomes asymmetric about zero.
As a result, the thermodynamic potential takes the form in the two-flavor case,
\begin{align}
 &\Omega(\mu,T,B;m,q) = \frac{m^2}{4G} + N_c\sum_{f=u,d} \Omega_f,
\end{align}
where,
\begin{align}
 &\Omega_f = -\frac{|e_fB|T}{4\pi} \int \frac{dp_z}{2\pi} \sum_k \bigg\{ \sum_{n,\zeta,\epsilon} {\rm ln}\left[\omega_k^2 + (E^{f}_{p_zn\zeta\epsilon} - \mu)^2\right] \notag \\
 &~~~~~~~~~~~~~~~~~~~~~~~~~~+ \sum_{\epsilon} {\rm ln}\left[\omega_k^2 + (E_{p_z\epsilon} - \mu)^2\right] \bigg\},
% &\Omega' = -\frac{1}{2} \frac{|e_fB|}{(2 \pi)^{2}} \notag \\
% &\times \int dp_z \bigg\{\sum_{n\zeta\epsilon} \left[ | E_{p_zn\zeta\epsilon}^{\rm hLL} - \mu | + \frac{2}{\beta} \ln \left( 1 + e^{-\beta | E_{p_zn\zeta\epsilon}^{\rm hLL} - \mu |} \right) \right] \notag \\
% &~~~+\sum_{\epsilon} \left[ | E_{p_z\epsilon}^{\rm LLL} - \mu | + \frac{2}{\beta} \ln \left( 1 + e^{-\beta | E_{p_z\epsilon}^{\rm LLL} - \mu |} \right) \right] \bigg\}.
\end{align}
%where $\Omega'$ represents the thermodynamic potential in the one-flavor case. \par
with the Matsubara frequency, $\omega_k=(2k+1)\pi T$.
To investigate the response of quark matter to the weak magnetic field $B$, the thermodynamic potential is expanded about $B$,
\begin{align}
 &\Omega(\mu,T,B\,;m,q) = \notag \\
 &~~~\Omega^{(0)}(\mu,T\,;m,q)+eB\,\Omega^{(1)}(\mu,T\,;m,q)+\cdots, \label{thermo}
\end{align}
where $e$ denotes the elementary charge.
Since the vacuum part in $\Omega^{(0)}$ has divergence, it must be regularized by ,e.g., the proper time regularization (PTR) \cite{nakano}.
LLL contributes only to $\Omega^{(1)}$ because the energy spectrum does not depend on $B$ and 
the $B$ dependence only emerges through the Landau degeneracy factor.
On the other hand, hLLs contribute to the all order terms of $B$.

\begin{figure*}[ht]
 \centering
 \begin{tabular}{c}
      \begin{minipage}{0.5\hsize}
        \begin{center}
          \includegraphics[width=5.5cm, angle=270]{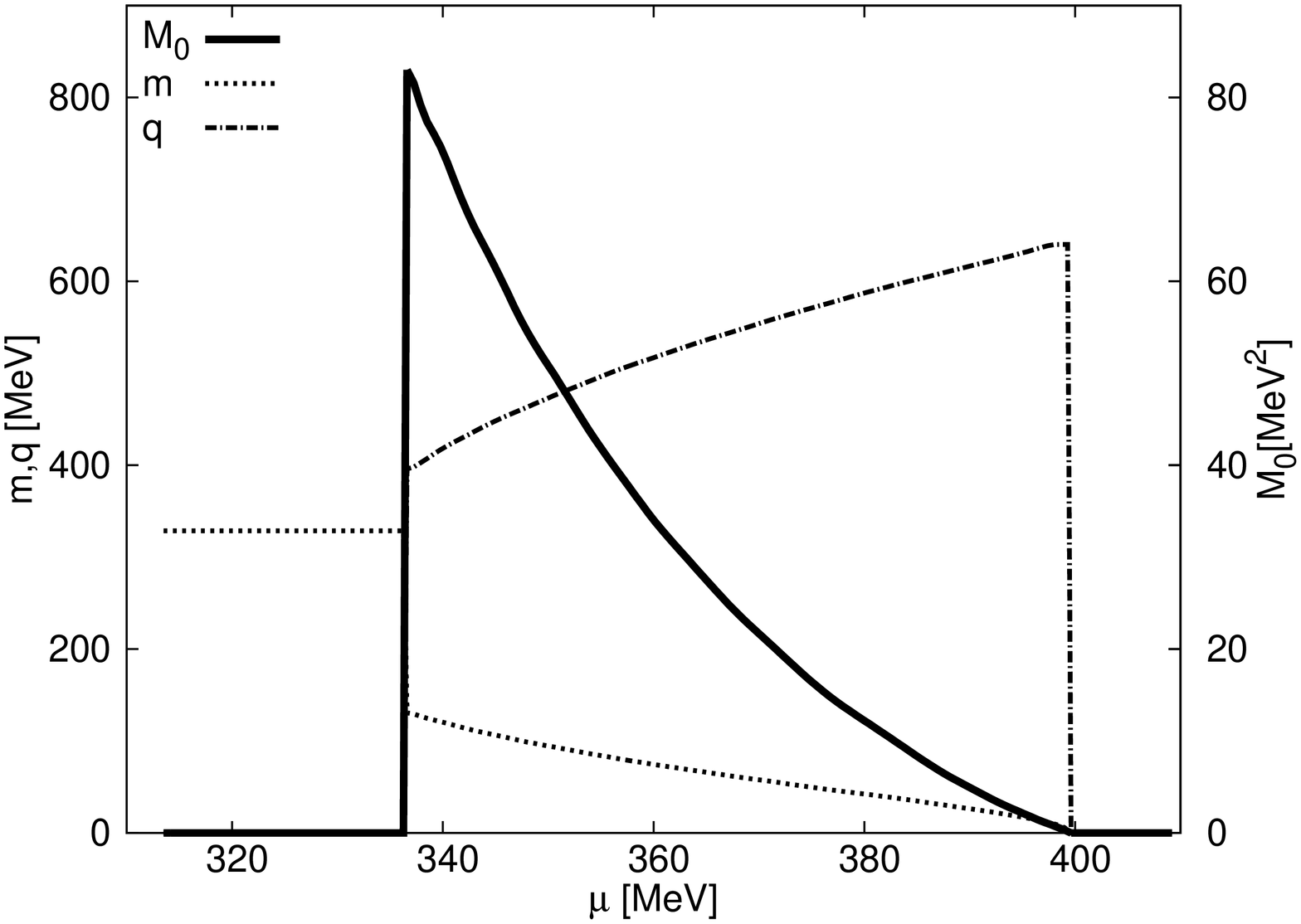}
          \vspace{0.7cm}
          \hspace{1.6cm} $(a) T=0$
        \end{center}
      \end{minipage}

      \begin{minipage}{0.5\hsize}
        \begin{center}
          \includegraphics[width=5.5cm, angle=270]{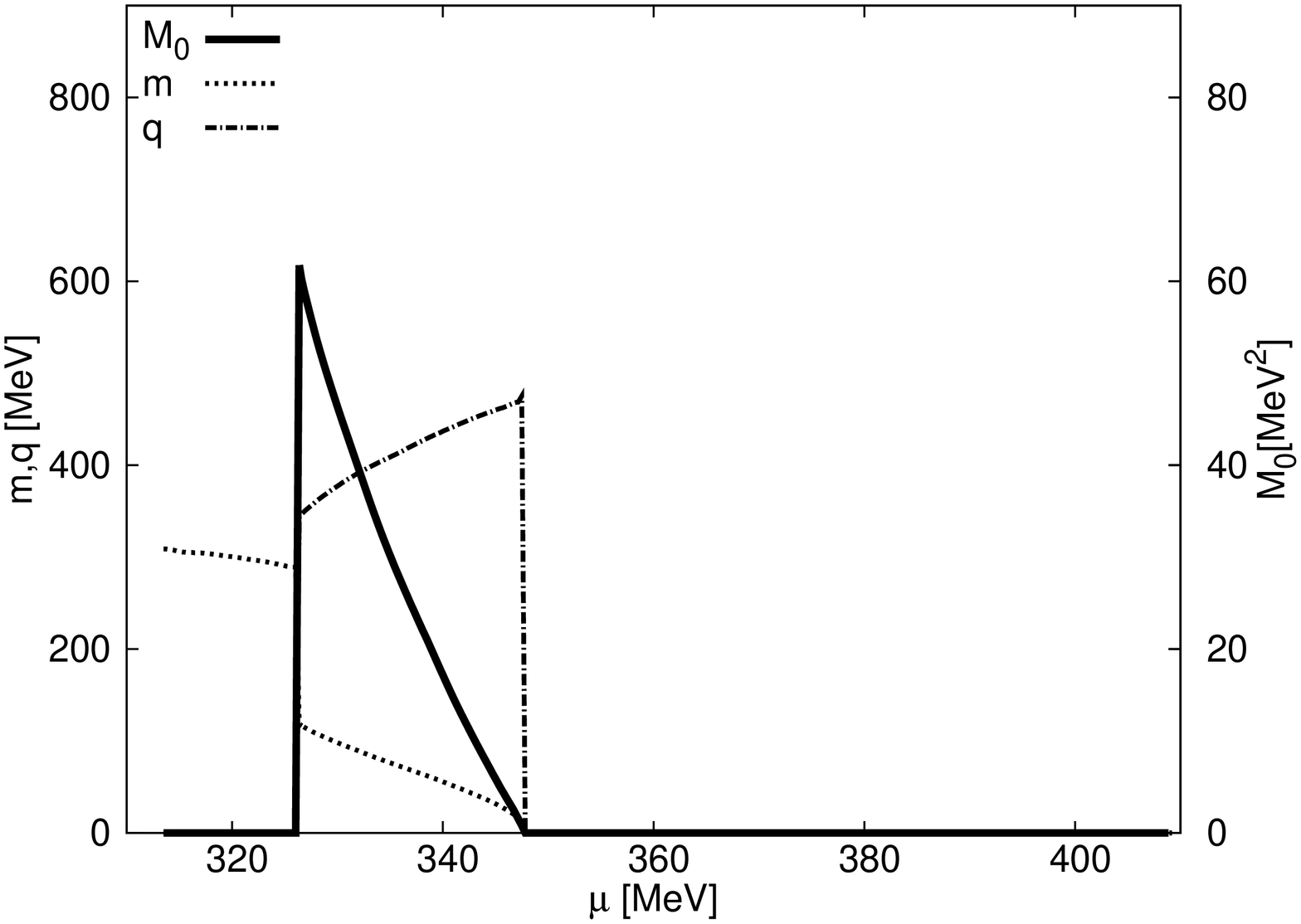}
          \vspace{0.7cm}
          \hspace{1.6cm} (b) $T\simeq30{\rm MeV}$
        \end{center}
      \end{minipage}      
 \end{tabular}
 \caption{The chemical potential dependence of the order parameters and spontaneous magnetization at $B=0$}
 \label{order}
\end{figure*}

The magnetization can be deduced from the thermodynamic potential as, 
\begin{align}
 M = -\frac{\partial \Omega_{\rm min}(\mu,T,B)}{\partial B}, \label{mag}
\end{align}
where $\Omega_{\rm min}$ represents the minimized thermodynamic potential about the order parameters and only depends on $\mu,T$ and $B$.
%Then, if magnetization survives in the limit, $B\rightarrow 0$, we can say that there is spontaneous magnetization.
%Since the order parameters satisfy the minimum conditions, $\partial \Omega / \partial m = 0,\partial \Omega %/ \partial q = 0$,
%they can be also expanded with respect to $B$ in a tiny magnetic field.
%begin{align}
%m(\mu,T,B) &= m^{(0)}(\mu,T)+eBm^{(1)}(\mu,T)+\cdots, \notag \\
%q(\mu,T,B) &= q^{(0)}(\mu,T)+eBq^{(1)}(\mu,T)+\cdots,
%\end{align}
% and others, $m^{(i)},q^{(i)}$, represent the correction by $B$.
Taking the limit, $B\rightarrow 0$, we find the spontaneous magnetization in the form,
\begin{align}
 M_{0} = -e\Omega^{(1)}(\mu,T;m=m^{(0)},q=q^{(0)}), \label{mag}
\end{align}
where $m^{(0)}$ and $q^{(0)}$ represent the minimal values for $\Omega^{(0)}$.
In the following, we will figure out the peculiar role of LLL and show that it leads to spontaneous magnetization.

%For obtaining the correct result, we must take account of the anomalous contribution caused by spectral asymmetry, which is similar to the argument of particle number.
For the evaluation of $\Omega^{(1)}$, we must carefully treat the effect of chiral anomaly.
%{\bf It has been shown that LLL induces anomalous quark number in the DCDW phase \cite{tatsumi}.}
According to Refs. \cite{niemi,niemi2}, spectral asymmetry generally gives rise to anomalous particle number, 
\begin{align}
 N_{{\rm anom}}= -\lim_{s \to +0}\frac{1}{2}\sum_k {\rm sign}(\lambda_k)|\lambda_k|^{-s}, \label{anom}
\end{align}
where $\lambda_k$ is the eigenvalue of the arbitrary Dirac Hamiltonian.
Spectral asymmetry is ill-defined as it is and needs a proper regularization without violating the gauge invariance.
%Here, it is regularized on the energy which does not violate the gauge invariance since it is ill-defined without some regularization.
In the DCDW phase, LLL exhibits spectral asymmetry to induce anomalous particle number  proportional to $B$ \cite{tatsumi}.
Then, the LLL contribution in $\Omega^{(1)}$ can be decomposed into three terms,
\begin{align}
 \Omega^{(1),{\rm LLL}} = \Omega^{(1),{\rm LLL}}_{\rm vac} + \Omega^{(1),{\rm LLL}}_{\mu} + \Omega^{(1),{\rm LLL}}_T, \label{lll}
\end{align}
where
\begin{align}
 \Omega^{(1),{\rm LLL}}_{\rm vac} =& -\frac{N_c}{4\pi} \int \frac{dp_z}{2\pi} \sum_{\epsilon} \left|\omega_\epsilon\right|, \\
 \Omega^{(1),{\rm LLL}}_{\mu} =& -\frac{N_c}{2\pi} \int \frac{dp_z}{2\pi} \sum_{\epsilon} \left( \mu - \omega_\epsilon \right) \theta(\omega_\epsilon)\theta(\mu - \omega_\epsilon) \notag \\
                                           & + \frac{\mu N_c}{4\pi} \eta_H, \\
 \Omega^{(1),{\rm LLL}}_T =& -\frac{N_cT}{2\pi} \int \frac{dp_z}{2\pi} \sum_{\epsilon} \ln\left(1 + e^{-\beta|\omega_\epsilon - \mu|} \right),
\end{align}
with $\omega_{\epsilon} = \epsilon\sqrt{p_z^2 + m^2} + q/2$.
The density dependent term $\Omega_\mu^{(1), {\rm LLL}}$ includes the anomalous contribution, $\frac{\mu N_c}{4\pi}\eta_H$, caused by spectral asymmetry.
The $\eta$-invariant, $\eta_H$, renders
\begin{align}
 \eta_H &\equiv \lim_{s \to +0}\int \frac{dp_z}{2\pi} \sum_{\epsilon}|\omega_\epsilon|^{-s}{\rm sign}(\omega_\epsilon) \notag \\
           &= \left\{ \begin{array}{lc} 
                         -\frac{q}{\pi} & (m>q/2) \\
                         -\frac{q}{\pi} + \frac{2}{\pi}\sqrt{q^2/4 - m^2} & (m<q/2)
                         \end{array} \right..
\end{align}
When $m>q/2$, this quantity agrees with the contribution of the chiral anomaly represented by the Wess-Zumino-Witten (WZW) term \cite{son}.
The WZW term does not depend on $m$ but it vanishes in the limit, $m \rightarrow 0$.
The contribution of hLLs to $\Omega^{(1)}$ should be carefully evaluated by expanding the thermodynamic potential with respect to $B$ after the summation over $n$.
Then, the hLLs contribution in $\Omega^{(1)}$ can be similarly decomposed into three terms,
\begin{align}
 \Omega^{(1),{\rm hLL}} = \Omega^{(1),{\rm hLL}}_{\rm vac} + \Omega^{(1),{\rm hLL}}_{\mu} + \Omega^{(1),{\rm hLL}}_T,
\label{hll}
\end{align}
which does not include the anomalous contribution since hLLs have no spectral asymmetry.  We find  that the three terms are the even function of $q$, and $\Omega^{(1),{\rm hLL}}_{\rm vac} =- \Omega^{(1),{\rm LLL}}_{\rm vac}$. Thus  {\bf $\Omega^{(1)} = \Omega^{(1),{\rm LLL}} + \Omega^{(1), {\rm hLL}}$ }
does not diverge without any regularization and renders
\begin{eqnarray}
\Omega^{(1)}&=&\frac{\mu N_c}{4\pi} \eta_H\nonumber\\
&-&\frac{N_c}{4\pi} \int \frac{dp_z}{2\pi} \sum_{\epsilon} \sum_{\tau=\pm1} \tau\left( \mu - \tau\omega_\epsilon \right) \theta(\tau\omega_\epsilon)\theta(\mu - \tau\omega_\epsilon)\nonumber\\
&-& \frac{N_cT}{4\pi} \int \frac{dp_z}{2\pi} \sum_{\epsilon}\sum_{\tau=\pm1} \tau\ln\left(1 + e^{-\beta|\omega_\epsilon - \tau\mu|} \right). \label{ome1}
\end{eqnarray}
The first term can be interpreted as the contribution of anomaly and the second and third terms as the contribution of positive-energy valence quarks.
Note that the even function of $q$ in Eq.~(\ref{lll}) is  completely canceled by the corresponding one in Eq.~(\ref{hll}) to make $\Omega^{(1)}$ the odd function of $q$. It vanishes in the limit, $m\rightarrow 0$, which behavior 
may be physically reasonable
because there should be no condensate and $q$ should be redundant in this limit.
% and if $\Omega^{(1)}$ does not vanish in this limit, the residual $\Omega^{(1)}$ %necessarily depends on $q$.
Consequently we can see that $\Omega^{(1)}$ emerges only when $m\neq 0, q\neq 0$.
In other words, quark matter has the spontaneous magnetization only in the DCDW phase.

Note that the spontaneous magnetization takes a special form for some peculiar values of the parameters, $m, q$, as well as $\mu, T$.
At $\mu<m-q/2$ and $T=0$, where is no valence quark,
the magnetization only comes from the WZW term, $M_0 = \frac{\mu N_cq}{4\pi^2}$, argued in Ref. \cite{son}.
Unfortunately, such a situation is not realized in the present calculation
because there always exist valence quarks in the DCDW phase, which is given rise to by the nesting effect of the Fermi surface \cite{nakano,peierls,gruner}.
On the other hand,
when $m$ becomes small compared to $\mu$ or $T$ but $q$ still is not small, $M_0$ is evaluated to be
\begin{align}
 M_0 &= -e\frac{N_c}{8\pi^2}\sum_{\sigma=\pm1}\sigma\,{\rm Re}\psi\left(\frac{1}{2} + i\frac{q/2 - \sigma\mu}{2\pi T}\right)m^2 \notag \\
       &~~~~~~~~~+ {\cal O}\left(m^4\right),
\end{align}
where $\psi$ is the digamma function.
Then, the leading order about $m$ is $m^2$ by the symmetry, $m\rightarrow -m$.
%Then, the leading order about $m$ is $m^2$ for the $Z_2$ symmetry of $m$.

%Generally, magnetization is evaluated by considering the response to the external magnetic field,

%\begin{figure*}[ht]
% \centering
% \begin{tabular}{c}
%      \begin{minipage}{0.5\hsize}
%        \begin{center}
%          \includegraphics[width=5cm, angle=270]{mag0tem.ps}
%          \vspace{0.7cm}
%          \hspace{1.6cm} (a) $T=0$
%        \end{center}
%      \end{minipage}

%      \begin{minipage}{0.5\hsize}
%        \begin{center}
%          \includegraphics[width=5cm, angle=270]{mag005.ps}
%          \vspace{0.7cm}
%          \hspace{1.6cm} (b) $T\simeq30{\rm MeV}$
%        \end{center}
%      \end{minipage}      
% \end{tabular}
% \caption{The chemical potential dependence of spontaneous magnetization}
% \label{magnet}
%\end{figure*}
The magnetization is numerically evaluated at $T=0$ and $T\simeq 30{\rm MeV}$, for example. 
In this calculation, we choose $\Lambda = 660{\rm MeV}$ as the cutoff in PTR and $G\Lambda^2 = 6.35$, which reproduce pion decay constant $f_{\pi} = 93{\rm MeV}$ and constituent quark mass $\simeq 330{\rm MeV}$ in the vacuum.
FIG. \ref{order} shows the chemical potential dependence of the order parameters and the spontaneous magnetization {\bf $M_0$}.
There are three phases, the DCDW phase ($m\neq 0$, $q \neq 0$), the homogeneously chiral-broken phase ($m\neq 0$, $q=0$) and the chiral-restored phase ($m=0$).
We can see that {\bf $M_0$} becomes nonzero only in the DCDW phase and has discontinuity at the transition point from the homogeneously chiral-broken phase. At this point, the order parameters also discontinuous, which implies the phase transition is of the first order.
On the other hand, near another transition point to the chiral-restored phase, the spontaneous magnetization is proportional to $(m^{(0)})^2$.
Furthermore, as temperature increases, the spontaneous magnetization decreases and the region of magnetized phase gets narrower.

Considering a sphere of the uniform magnetization $M_0$, the magnetic field, $B=\frac{8\pi}{3}M_0$ is produced on the surface. 
For the sphere of quark matter with a constant density in the DCDW phase at zero temperature,
the magnetic field is estimated $B \sim 10^{16}{\rm G}$ on the surface, which may be comparable with the observation of magnetars.
However, this estimate might be too rough because density is not constant in neutron stars. Finite shell structure of the DCDW phase may be another possible realization.
Considering that the DCDW phase gets narrow at high temperature,
magnetic field may become much lower than this estimate in hot neutron stars.
Actually the spontaneous magnetization vanishes with the disappearance of the DCDW phase at $T\sim 100 {\rm MeV}$.
At $T\simeq 30{\rm MeV}$, which may be a relevant temperature in newly born neutron stars, the magnetic field remains to be a similar order of magnitude.
%At this point, the continuity of $m$ implies the second order phase transition.
%At a glance, $q$ looks like discontinuous at this point, but 
%the thermodynamic potential does not depend on $q$ in the chiral-restored phase and $q$ is redundant.

%In the previous paper \cite{nakano}, it has been suggested that spin density wave is realized in the DCDW phase, instead of spontaneous magnetization. 

%In the system of neutrons, the external magnetic field couples with the spin of neutrons only through the anomalous magnetic moment because neutrons does not have the electric charge.
%So, the expectation value of spin can be regarded as the magnetization \cite{eto,hashimoto}.
In Ref. \cite{nakano}, the expectation value of the magnetic moment, $\langle\bar{\psi}({\bf r})\sigma^{12}\psi(\bf r)\rangle$, is evaluated, which behaves like spin density wave and  vanishes after the spatial average.
This seems to contradict with the present results. However, we can see that a careful application of the Gordon identity leads to the present results.
The external magnetic field minimally couples with the quark field through the covariant derivative, $\int d^4x\,\bar{\psi}Q\gamma^{\mu}\psi A_{\mu}$, in the Lagrangian,
where $Q$ is the electric charge matrix in the flavor space.
According to Ref. \cite{nielsen}, this term can be decomposed into the form,
\begin{align}
 \int d^4x\,\bar{\psi}Q\gamma^{\mu}\psi A_{\mu} &= \frac{B}{2m} \int d^4x \bigg[ \bar{\chi}Q\sigma^{21}\chi + \bar{\chi}Q2ixD_2\chi \notag \\
&~~~~~~~~~~~~~~+ \bar{\chi}Q\sigma^{23}i\gamma^5\tau_3q\,x\chi \bigg], \label{gordon}
\end{align}
 where the Landau gauge is taken, ${\bf A}=(0,Bx,0)$,
and $\chi$ represents the quark field after the Weinberg transformation, $\chi = e^{i\gamma^5\tau_3qz/2}\psi$.
Here we have used the modified Gordon decomposition by using the Dirac equation in the presence of DCDW, instead of the usual treatment by the free Dirac equation.
Consequently, we have the expectation value of the magnetic moment, $\langle\bar{\chi}(\bf{r})\sigma^{12}\chi(\bf{r})\rangle$, instead of the above one in terms of $\psi$,  which never vanishes after the spatial average.
There appear two contributions in (\ref{gordon}) besides the magnetic moment: 
the second term  can be interpreted as the angular momentum and the third term proportional to $q$ comes from the operator inherent in the DCDW phase.
Thus, the magnetization discussed here should be regarded as the statistical average of these operators.

We have shown that quark matter has spontaneous magnetization in the DCDW phase
where the response to the external magnetic field changes from the one in the homogeneously chiral-broken phase or the chiral-restored phase, that is, $\Omega^{(1)}$ emerges in the thermodynamic potential.
In Ref. \cite{son}, they have simply discussed the spontaneous magnetization by using the WZW term in the effective Lagrangian written in terms of only mesons.
However, we have seen that $\Omega^{(1)}$ has not only the anomalous contribution closely related to the WZW term
but also the contribution of valence quarks.
In this sense, the fermion degrees of freedom  and their dynamics are indispensable.
Thus we can say that DCDW provides a realistic ground where the spontaneous magnetization is realized.
%Therefore, 
%we can evaluate spontaneous magnetization including the contribution of quark matter.

The spontaneous magnetization discussed here has an interesting feature different from the usual spin alignment \cite{tatsumi2}: 
%urthermore, the possibility of spontaneous magnetization by the spin alignment has been discussed in quark gas 
the magnetization in the DCDW phase is caused by the different operators from the naive magnetic moment.
Further discussions about the peculiar magnetic properties such as magnetic susceptibility or the Nambu-Goldstone mode (spin wave) will be presented in another paper.

We have emphasized that the spectral asymmetry is important for the mechanism of the spontaneous magnetization.
%Assuming $\varphi({\bf r})$ is a eigenfunction of the energy, $H\varphi({\bf r}) = E(\Delta({\bf r}))\varphi({\bf r})$, $\varphi'=\gamma_0\gamma_5\varphi$ is the eigenfunction with the eigenvalue $-E(\Delta^*({\bf r}))$ \cite{tatsumi}.
Since the complex order parameter $\Delta({\bf r})$ is necessary for the energy spectrum to be asymmetric \cite{tatsumi},
it is conceivable that there is no spontaneous magnetization in the ``real kink crystal'' \cite{basar,basar2} phase with the real order parameter $\Delta({\bf r})$.
On the other hand, the spontaneous magnetization should emerge in the similar way in the phase with hybrid chiral condensate \cite{nishiyama}.
 
It has been discussed that the NJL model, in the mean field approximation, exhibits magnetic catalysis \cite{suganuma,klevansky,gusynin}, while the recent lattice calculation shows inverse magnetic catalysis \cite{bali,endrodi}.
To explain this phenomenon an idea of the effective four-Fermi coupling has been proposed within the NJL model \cite{braun,mueller},
 where the coupling strength depends on the external magnetic field by considering the coupling of the magnetic field with the quark and gluon loops constituting the four-Fermi coupling. 
Even if this is the case, however, the response to the tiny external magnetic field may be little changed and our findings should not be affected.

In order to explain the strong magnetic field in neutron stars by the present mechanism, it is necessary to study the appearance of the DCDW phase in more realistic conditions such as the charge neutrality or the chemical equilibrium.
It is also an open question that the DCDW phase survives when quarks have the finite current mass \cite{karasawa}.

The authors thank T.-G. Lee for useful discussions.
This work is partially supported by Grants-in-Aid for Japan Society for the Promotion of Science (JSPS) fellows No. 27-1814 
and Grants-in-Aid for Scientific Research on Innovative Areas 
through No. 24105008 provided by MEXT.

\bibliographystyle{elsarticle-num}
\bibliography{reference}

%\begin{thebibliography}{99}
% \bibitem{nakano} E. Nakano, T. Tatsumi, Phys. Rev. D 71, 114006 (2005)
% \bibitem{nickel} D. Nickel, Phys. Rev. D 80, 074025 (2009)
% \bibitem{muller} D. Muller, M. Buballa, J. Wambach, Phys. Lett. B 727, 240 (2013)
% \bibitem{basar} G. Basar, G. V. Dunne, Phys. Rev. D 78, 065022 (2008)
% \bibitem{tatsumi} T. Tatsumi, K. Nishiyama, S. Karasawa, Phys. Lett. B, to be published (2015)
% \bibitem{skokov} V. V. Skokov, A. Yu. Illarinov, V. D. Toneev, Int. J. Mod. Phys. A 24, 5925 (2009)
% \bibitem{olausen} S. A. Olausen, V. M. Kaspi, arXiv:1309.4167v3, (2013)
% \bibitem{frolov} I. E. Frolov, V. Ch. Zhukovsky, K.G. Klimenko, Phys. Rev. D 82, 076002 (2010)
% \bibitem{niemi} A. J. Niemi, G. W. Semenoff, Phys. Rep. 135(3), 99 (1986)
% \bibitem{son} D. T. Son, M. A. Stephanov, Phys. Rev. D 77, 014021 (2008)
% \bibitem{ggl} D. Nickel, Phys. Rev. Lett. 103, 072301 (2009)
%\end{thebibliography}

\end{document}